\documentclass[a4paper]{article}
\usepackage[a4paper]{geometry}
\usepackage{amsmath,amsfonts,amsthm} 
\usepackage{fancyhdr}
\usepackage{algorithm2e,setspace}
\usepackage{graphicx}
\usepackage[font=scriptsize,labelfont=bf]{subcaption}
\usepackage[font=small,labelfont=bf]{caption}
\usepackage{tabularx}
\usepackage{amsmath}
\usepackage{epstopdf}
\usepackage{booktabs}
\usepackage{multirow}
\usepackage{longtable}
\usepackage[numbers,sort&compress]{natbib}
\newgeometry{left=2cm,bottom=2.5cm,right=2cm,top=2.5cm}

\graphicspath {{Figures/}}
\title{
		\huge Micro-scale process modeling and residual stress prediction in fiber-reinforced polymers using refined structural models \\ 
}

\author{M. H. Nagaraj\footnote{Postdoctoral Research Associate, e-mail: manish\_nagaraj@uml.edu}, M. Maiaru\footnote{Associate Professor, e-mail: marianna\_maiaru@uml.edu}    \\
     Department of Mechanical Engineering, \\
     University of Massachusetts Lowell, MA 01854, United States} 

\date{}

\begin{document}

\maketitle
\vspace*{3.cm}
\parbox{12cm}{\large Manuscript 
\large }

\vspace*{2.cm} \baselineskip=6.2mm \vspace*{2.cm} \vspace*{4.cm}
\parbox{12cm}{{\em Corresponding Author:\\}
     M. Maiaru, \newline
     Department of Mechanical Engineering, \newline
     University of Massachusetts Lowell, \newline
     Lowell 01854,\newline
     MA, United States \newline
     e-mail: marianna\_maiaru@uml.edu
     }

\newpage

\doublespacing

$\;$\vspace*{20mm}

\begin{center}
\section*{\em Abstract}
\end{center}
\em
\doublespacing
The present work introduces a novel numerical approach for the process modeling of fiber-reinforced thermoset polymers at the micro-scale level, that can be used to predict curing-induced residual stresses. The cure kinetics is described using an auto-catalytic phenomenological model and an instantaneous linear-elastic constitutive law is used to evaluate the stress state evolution as a function of the degree of cure and time. The proposed method is based on refined structural theories derived from the Carrera Unified Formulation (CUF). A series of numerical assessments is carried out to evaluate the performance of CUF models in micro-scale curing analysis -- considering neat resin, a single-fiber repeating unit cell, and a representative volume element with 20 randomly distributed fibers. Comparing the CUF predictions with reference 3D finite element (3D-FE) models demonstrates the accuracy of the present approach in stress analysis. It is also shown that CUF models are an order-of-magnitude faster than those based on conventional 3D-FE, for similar accuracy of results.
\rm
\\
\\
\textbf{Keywords}: Process modeling, CUF, higher-order structural modeling, curing, thermosets  

\newpage
\section{Introduction}

Fiber-reinforced plastics (FRP) have become increasingly popular in many industries, including aerospace, automotive, and wind, due to their outstanding mechanical properties, such as specific strength and stiffness. However, their complex mechanical response and uncertainties in their properties due to manufacturing imperfections lead to the use of larger margins of safety and an overly conservative design, which restricts the utilization of the full potential of composite materials and limits structural efficiency. A further bottleneck is the large design space associated with composite structures, which renders physical testing a lengthy and resource-intensive task.

The manufacturing process of thermoset FRP has a significant effect on both the final geometry of the part as well as its mechanical performance. The manufacturing process, or curing, consists of an exothermic chemical reaction, during which material properties change as a function of time and temperature. The thermal expansion mismatch between fibers and matrix leads to differential expansion within the composite microstructure. Such thermal expansion mismatch at the level of the constituents, in combination with chemical shrinkage of the matrix and the thermo-mechanical properties evolution during the cure cycle, results in self-equilibrated residual stresses \cite{baran2017review, kim1989residual}. High residual stress levels during the cure cycle can lead to the formation of micro-cracks within the matrix, significantly reducing the composite mechanical performance and service life \cite{chekanov1995cure,hu2018investigation}. The robust design and optimization of composite structures thus requires an accurate understanding of the development of residual stresses within the manufactured part.

Computationally efficient multiscale process modeling is needed to predict curing-induced residual stresses accurately \cite{MS_curing_1}. Different numerical techniques can reproduce the relevant fundamental physics across the relevant scales in fiber-reinforced composites, including the evolution of the mechanical and non-mechanical properties of the matrix as a function of curing, local fiber constraint of the curing matrix, thermal gradients, and stress concentrations induced by complex tow architectures \cite{MS_curing_2,MS_curing_3}. Recent work proved that it is possible to virtually reproduce the crosslinking formation of the polymer during curing at the nano-scale using Molecular Dynamics (MD) simulations \cite{polymer_MD_1,polymer_MD_2}. MD simulations accurately represent the chemical composition of the resin and its curing agent, and predict the mechanical property evolution of the matrix as a function of the crosslinking density \cite{polymer_MD_3,polymer_MD_4}. The local variation in fiber volume fraction driven by the stochastic distribution of the fibers induces local variability in the residual stress state, which in turn affects the mechanical properties of the curing resin. Additionally, the proximity of neighboring fibers can act as stress raisers in the microstructure that triggers premature failure \cite{fiber_variation_1,fiber_variation_2,maiaru2018characterization}. Modeling the micro-scale resolution enables the explicit representation of the reinforcing fibers and the resulting stochastic property variabilities within the composite needed to represent a realistic structure. Thus, the explicit modeling of the fibers at the micro-scale is crucial to predict the composite failure behavior accurately \cite{RVE_modelling_1,RVE_modelling_2,RVE_modelling_3,RVE_modelling_4,RVE_modelling_5}. While traditional 3D Finite Element (3D-FE) analysis is a preferred tool to analyze complex structures at the macro-scale \cite{macro_FEA_1,curing_stress_3}, computational micromechanical models based on conventional FE can incur prohibitively high costs, especially when they are used in a multiscale setting to inform composite material behavior in a structural-level analysis \cite{micromech_cost_1}. Thus, a computationally-efficient numerical approach is necessary to model the microstructure, which would eventually form part of multiscale process modeling frameworks. 

Several cure kinetics models have been proposed in the literature \cite{cure_kinetics_2,cure_kinetics_3,cure_kinetics_4}, with one of the most popular formulations being the phenomenological model by Kamal and Sourour \cite{kamal1973_cure_kinetics}. These kinetic models have been used in various numerical investigations on the prediction of residual stresses during the curing process, as well as the influence of these stresses on the effective mechanical properties of the composite. For instance, Ding et al. proposed a 3D thermo-viscoelastic model to simulate residual stresses in composite laminates during curing \cite{curing_stress_2}. Maiaru et al. investigated the influence of the manufacturing process on the transverse strength of unidirectional FRP using traditional FEs \cite{maiaru2018characterization}. More recently, Hui et al. developed a micro-scale viscoplastic model to investigate the effect of curing-induced stresses on the compressive strength of unidirectional FRP \cite{curing_stress_4}. Similarly, D’Mello et al. presented an approach to simulate the curing process of homogenized textile composites and subsequently evaluated the effect of the developed residual stresses on the tensile strength of the composite \cite{curing_stress_3}. 

This paper presents a novel computational framework based on an instantaneous linear elastic micro-scale curing model, developed using higher-order finite elements, that will enable the multiscale modeling of fiber-reinforced composites. The numerical modeling approach is based on the Carrera Unified Formulation (CUF), which is a generalized framework for developing higher-order structural theories. CUF-based models are capable of an accuracy approaching that of 3D-FEs at significantly reduced computational effort \cite{CUF_book}. The advantages of CUF have been demonstrated in recent years for various classes of problems, such as progressive damage and impact \cite{CUF_tensiledamage,CUF_compressivedamage,CUF_concrete_nagaraj,CUF_impact}, micromechanical analysis \cite{CUF_micro_1,CUF_micro_2,CUF_micro_3,CUF_micro_voids}, and the analysis of process-induced deformations in cured composite parts \cite{CUF_process_composite}. The present work combines a process modeling formulation for thermoset epoxies \cite{maiaru2018characterization, Shah_maiaru_polymers2021} with CUF theories to enable the micro-scale curing analysis of thermoset composites.

This work is organized as follows: Section 2 describes the higher-order FE structural modeling approach within CUF and the process modeling formulation. A series of numerical assessments is presented in Section 3 to verify the proposed approach using traditional FEs. The main conclusions are summarized in Section 4.

\section{Numerical modeling} \label{section_num_mod}

\subsection{Structural modeling: Carrera Unified Formulation} \label{sec_cuf}

The Carrera Unified Formulation (CUF) is a generalized framework to derive higher-order structural theories, and in combination with the Finite Element Method (FEM), can be used to develop higher-order numerical models. Specifically, CUF allows for the kinematic enrichment of beam (1D-CUF) and plate/shell (2D-CUF) elements by the use of additional interpolation functions, resulting in numerical models that approach the accuracy of 3D-FEA at significantly reduced computational effort.

\begin{figure}[htbp]
	\centering
	\includegraphics[scale=0.30]{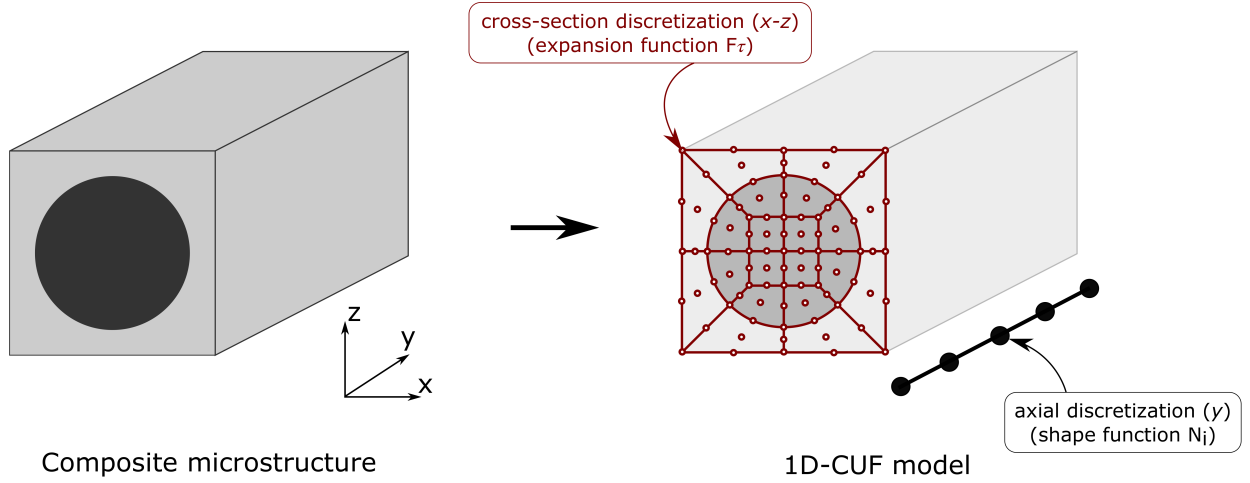}
	\caption{Modeling of a prismatic structure using 1D-CUF.}
	\label{CUF_1D_schematic}
\end{figure} 

The current work uses 1D-CUF models, wherein the cross-section of beam elements is explicitly defined by a set of additional interpolation functions, termed as expansion functions ($F_\tau$), as seen in Fig. \ref{CUF_1D_schematic}. In this approach, the displacement field \textbf{u} is defined as \cite{CUF_book}

\begin{equation}
\mathbf{u}(x,y,z) = F_\tau (x,z) \mathbf{u}_{\tau} (y),\ \tau = 1,2, ..., M
\end{equation} 

where M is the number of terms within the expansion function. Various classes of polynomial functions can be used as ($F_\tau$), and is chosen by the user. The most popular choice of expansion function are those based on Taylor series \cite{CUF_TE} and Lagrange polynomials \cite{CUF_LE}. Various other functions have been proposed to enhance the cross-sectional interpolation \cite{CUF_Legendre, CUF_misc_Ft}. The present work considers the use of Lagrange polynomials as $F_\tau$, which consist of nodal interpolation functions within the cross-sectional discretization, see Fig. \ref{CUF_1D_schematic}. This leads to a purely displacement-based formulation as seen below:

\begin{equation}
\begin{aligned}
u^x (x,y,z) = \sum_{i=1}^{Nnode} F_i (x,z)\cdot u_i^x (y) \\
u^y (x,y,z) = \sum_{i=1}^{Nnode} F_i (x,z)\cdot u_i^y (y)  \\
u^z (x,y,z) = \sum_{i=1}^{Nnode} F_i (x,z)\cdot u_i^z (y)  
\end{aligned}
\label{LE_disp_interp_eqn}
\end{equation}
 
where $u_i^x$, $u_i^y$, and $u_i^z$ are the translational degrees of freedom (DOF) of node \textit{i}. Furthermore, the use of cross-sectional Lagrange elements allows for the explicit modeling of each component domain within the structure, and is known as Component-Wise modeling \cite{CUF_CW, CUF_CW_2}.

\subsubsection*{Finite element formulation}

The stress and strain fields are defined using the Voigt notation as

\begin{equation}
\begin{aligned}
\boldsymbol{\sigma} = \{\sigma_{xx},\sigma_{yy},\sigma_{zz},\sigma_{xy},\sigma_{xz},\sigma_{yz}\}\\
\boldsymbol{\varepsilon} = \{ \varepsilon_{xx},\varepsilon_{yy},\varepsilon_{zz},\varepsilon_{xy},\varepsilon_{xz},\varepsilon_{yz} \}
\end{aligned}
\end{equation}

Considering infinitesimal strain theory, the displacement-strain relationship is described using the differential operator \textbf{D} as

\begin{equation}
\boldsymbol{\varepsilon} = \mathbf{Du}
\end{equation}

with
\[
\textbf{D}=
\begin{bmatrix}
\frac{\partial}{\partial x} & 0 & 0 \\
0 & \frac{\partial}{\partial y} & 0 \\
0 & 0 & \frac{\partial}{\partial z} \\
\frac{\partial}{\partial y} & \frac{\partial}{\partial x} & 0 \\
\frac{\partial}{\partial z} & 0 & \frac{\partial}{\partial x} \\
0 & \frac{\partial}{\partial z} & \frac{\partial}{\partial y} \\
\end{bmatrix}
\]

The stress-strain relation is given as

\begin{equation} \label{constitutive_law}
\boldsymbol{\sigma}(t) = \textbf{C}(T,\phi(t))\ \boldsymbol{\varepsilon}(t) 
\end{equation}

where \textbf{C} is the $6\times6$ material stiffness matrix. For the case of thermoset polymers, \textbf{C} depends on the temperature and the cure state which is quantified by the degree of cure ($\phi$) as described in Section \ref{sec_thermoset_process_modeling}. Discretizing the structure, schematically shown in Fig. \ref{CUF_1D_schematic}, along its axis with 1D finite elements (using interpolation functions $N_i$), and refining the cross-sectional kinematics using expansion functions $F_{\tau}$, the 3D displacement can be written as

\begin{equation} \label{3d_disp_field}
\mathbf{u}(x,y,z) = F_{\tau}(x,z)N_i(y)\mathbf{u}_{\tau i}
\end{equation} 

According to the principle of virtual work

\begin{equation} 
\delta L_{int} = \delta L_{ext}
\end{equation}

where the virtual variation of the internal strain energy $\delta L_{int}$ is defined as
\begin{equation} \label{PVD_Lint}
\delta L_{int} = \int_{V}\delta\boldsymbol{\varepsilon}^T\colon\boldsymbol{\sigma}
\end{equation}

Combining Eqs. \ref{constitutive_law}, \ref{3d_disp_field} and \ref{PVD_Lint}, the stiffness matrix can be derived as

\begin{equation}
\delta L_{int} = \delta\mathbf{u}^T_{sj}\mathbf{k}_{ij\tau s}\mathbf{u}_{\tau i}
\end{equation}

with 

\begin{equation}
\mathbf{k}_{ij\tau s} = \int_{l}\int_{A}\mathbf{D}^T(N_i(y)F_{\tau}(x,z))\ \mathbf{C}(T,\phi(t))\ \mathbf{D}(N_j(y)F_{s}(x,z))\ dA\ dl
\end{equation}

The 3x3 matrix $\mathbf{k}_{ij\tau s}$ is the Fundamental Nucleus (FN), and its definition remains invariant with respect to any given combination of interpolation function $N_i$ and expansion function $F_{\tau}$. The element-level stiffness matrix can then be computed by assembling the fundamental nuclei associated with each combination of the nodal indices $\{i,j,\tau,s\}$. The numerical model used in the current work requires a temperature DOF, in addition to the three mechanical DOF, in order to simulate the thermoset curing process. The temperature DOF can be accounted for in the FN by considering a thermal term $k_\theta$ as follows \cite{CUF_thermoelastic}

\begin{equation}
k_\theta = \int_{l}\int_{A}\boldsymbol{\nabla}^T(N_iF_\tau)\kappa\boldsymbol{\nabla}(N_jF_s)\ dA\ dl
\end{equation}

where $\kappa$ is the material thermal conductivity. Considering an uncoupled temperature-displacement problem, the augmented FN is now a $4\times4$ matrix, defined as

\begin{equation}
k_{ij\tau s}^{u\theta} = 
\begin{bmatrix}
\mathbf{k}_{u} & \mathbf{0} \\
 \mathbf{0}           & \mathbf{k}_\theta   \\
\end{bmatrix}
\end{equation}

\subsection{Process Modeling: Thermoset Curing} \label{sec_thermoset_process_modeling}

The thermoset cure kinetics, for a given cure cycle, is governed by an auto-catalytic phenomenological semi-empirical kinetic model \cite{kamal1973_cure_kinetics}, as follows

\begin{equation}
\frac{d\phi}{dt} = \left[ A_1 exp \left( - \frac{\Delta E_1}{RT}\right) + A_2 exp \left( - \frac{\Delta E_2}{RT}\right)\phi ^n\right] (1- \phi ^m)
\label{cure_kinetics_eqn}
\end{equation}

where $\phi$ is the degree of cure, $R$ is the gas constant, $T$ is the cure temperature at time $t$, with the activation energies denoted by $\Delta E_1$ and $\Delta E_2$. The constants $A_1$ and $A_2$, and the exponents $m$ and $n$, are determined experimentally. The thermal state of the thermoset is a consequence of the prescribed cure temperature and the heat generated due to the exothermic nature of the curing process, and the resulting temperature distribution is evaluated using the Fourier heat transfer model as follows

\begin{equation}
\rho c_p \frac{dT}{dt} = \kappa_i \frac{d^2T}{dt^2} + \frac{dq}{dt},\ with\ \frac{dq}{dt} = \rho H_T \frac{d\phi}{dt}
\label{fourier_HT_eqn}
\end{equation}

where $\rho$ and $c_p$ are respectively the density and specific heat of the epoxy, $\kappa_i$ is the thermal conductivity, $q$ is the instantaneous exothermic heat generated during the curing process, and $H_T$ is the total heat of reaction.

\begin{figure}[htbp]
	\centering
	\includegraphics[scale=0.4]{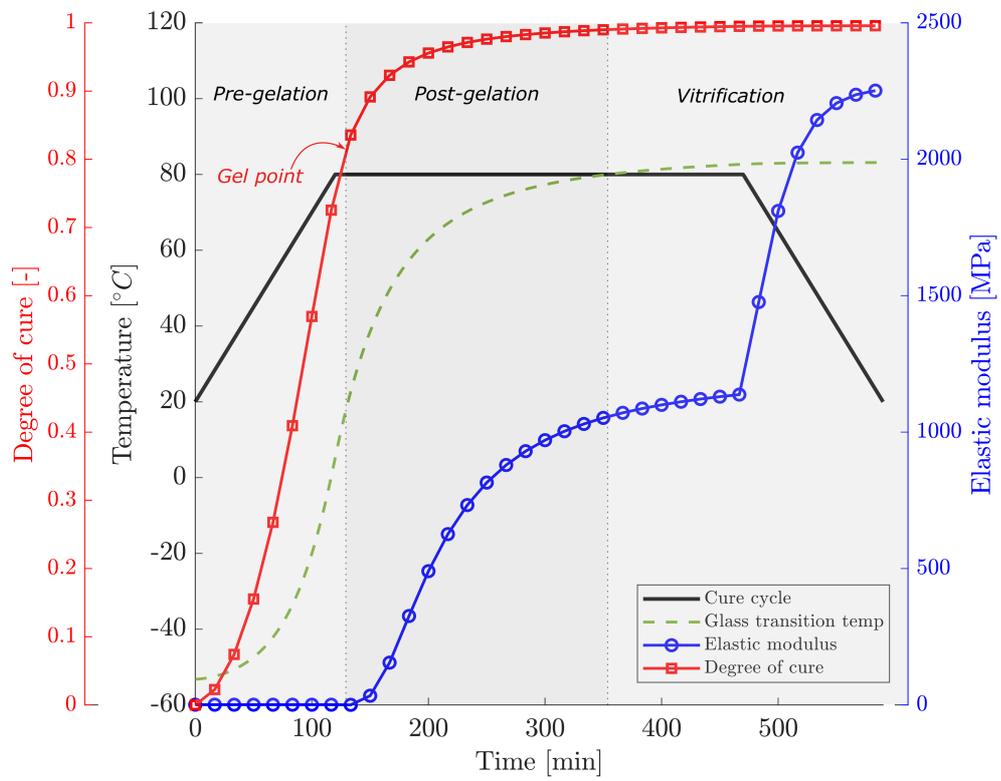}
	\caption{Temperature - degree of cure plot of RIM R135/H1366 epoxy resin system with evolution of the elastic modulus.}
	\label{RIMR_cure_cycle_plot}
\end{figure} 

During the curing process, the change in the degree of cure (as a function of time) results in an evolution of the chemo-rheological and thermo-mechanical properties of the thermoset. These properties have been previously characterized in-house for the RIM R135/H1366 epoxy resin system \cite{Shah_maiaru_polymers2021}, and their evolution for the manufacturer recommended cure cycle is plotted in Fig. \ref{RIMR_cure_cycle_plot}. The experimentally determined cure kinetics constants for this material system, required to evaluate Eq. \ref{cure_kinetics_eqn}, are listed in Table \ref{rimr_kinetics_constants}. The evaluated degree of cure, for a specific time $t$, can be used to determine the material state of the thermoset as seen in Fig. \ref{RIMR_cure_cycle_plot}. These properties can be used with an instantaneous linear-elastic constitutive model, previously described in Ref. \cite{maiaru2018characterization}, to predict the development of residual stresses ($\sigma_{i}$) as a function of the evolving thermal and chemical strains as follows

\begin{equation}
\sigma_i(t) =  \left[ C_{ij}(T,\phi(t)) \left[ \varepsilon_j^{tot}(t) - (\varepsilon_j^{therm}(T,\phi(t)) + \varepsilon_j^{shrink}(t)) \delta_j \right] \right], where\ \delta_j =  \begin{cases}
1 &\text{j = 1,2,3}\\
0 &\text{j $>$ 3}
\end{cases}
\label{instant_LE_stress}
\end{equation}

where $\varepsilon_j^{tot}(t)$, $\varepsilon_j^{therm}(t)$ and $\varepsilon_j^{shrink}(t)$ are respectively the total, thermal and chemical shrinkage strains, and $C_{ij}$ is the material stiffness, which evolves during the cure cycle as a function of temperature and degree of cure.

\begin{figure}[htbp]
	\centering
	\includegraphics[scale=0.45]{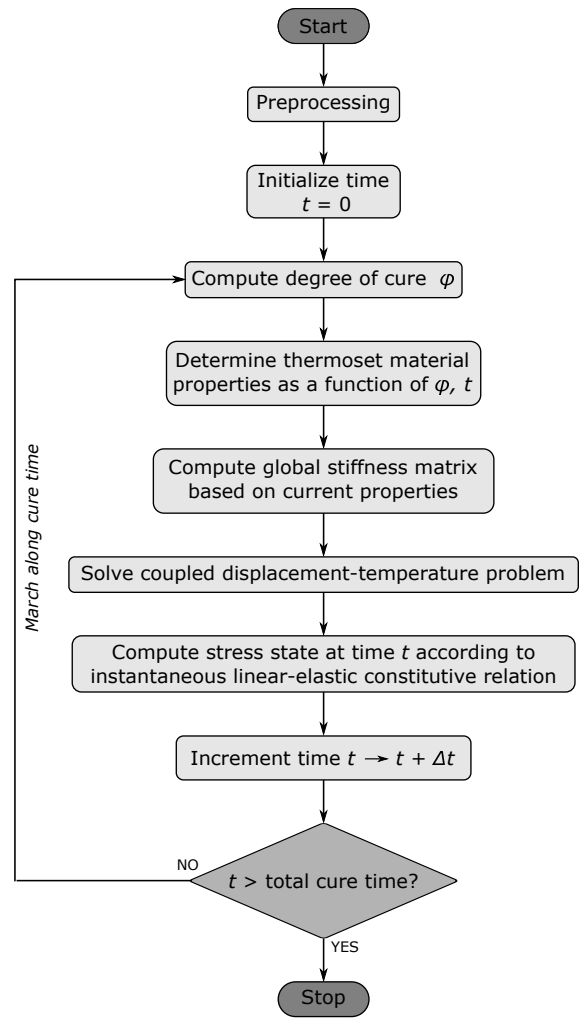}
	\caption{Schematic representation of the curing analysis.}
	\label{Curing_analysis_flowchart}
\end{figure}

The computational approach to simulate the curing process is developed based on the instantaneous linear-elastic nature of Eq. \ref{instant_LE_stress}. A time-based analysis, considering an incremental time period $\Delta t$, is performed over the cure cycle seen in Fig. \ref{RIMR_cure_cycle_plot}, and the degree of cure is evaluated at each time increment. The mechanical properties of the thermoset are determined as a function of the degree of cure, based on experimental characterization data. An  uncoupled displacement-temperature analysis, see Section \ref{sec_cuf}, is performed to compute the displacements and temperature fields. Finally, Eq. \ref{instant_LE_stress} is used to predict the residual stress developed within the thermoset material. A summary of the numerical analysis process is schematically presented in Fig. \ref{Curing_analysis_flowchart}.

\textbf{\begin{table}[htbp]
		\caption{Cure kinetics parameters for the RIM R135/H1366 epoxy resin system \cite{Shah_maiaru_polymers2021}. }
		\centering
		\begin{tabular}{lc}
			\toprule
		    Cure kinetic parameter      & Value             \\ \midrule
			Exponent $m$               &     0.4  $[-]$     \\ 
		    Exponent $n$               &     1.5  $[-]$     \\ 
			Rate constant $A_1$        &  3.6$\times 10^{9}\ [s^{-1}]$ \\ 
		    Rate constant $A_2$    & 0.01245 $[s^{-1}]$ \\ 
			Activation energy  $\Delta E_1$    &    85.3 $[kJ/mol]$ \\ 
		    Activation energy  $\Delta E_2$    &    11.1 $[kJ/mol]$ \\  \bottomrule
		\end{tabular} 
		\label{rimr_kinetics_constants}
	\end{table}
}  

\section{Numerical Assessments} \label{sec_results}

A series of numerical assessments is presented in this section, with the aim of verifying the proposed modeling approach, as well as to evaluate its performance with respect to conventional 3D-FEA. The micromechanical models of the fiber-reinforced epoxy are composed of IM7 fiber and RIM R135/H1366 epoxy resin, whose thermo-mechanical properties are listed in Table \ref{fiber_matprops} and \ref{RIMR_epoxy_matprops}, respectively. In each case, the cure simulation follows the cure cycle plotted in Fig. \ref{RIMR_cure_cycle_plot}.

\textbf{\begin{table}[htbp]
		\caption{Elastic and thermal material properties of the IM7 carbon fiber \cite{Shah_maiaru_polymers2021}. }
		\centering
		\begin{tabular}{lc}
			\toprule
			Material Property                                               & Value             \\ \midrule
			Density $\rho^f $                                               & 1780.0  $[kg/m^3]$ \\ 
			Axial modulus $E_{11}^f$                                        &  276.0  $[GPa]$    \\
			Transverse modulus $E_{22}^f$, $E_{33}^f$                       &   19.5  $[GPa]$    \\
			In-plane Poisson's ratio $\nu_{12}^f$, $\nu_{13}^f$             & 0.28 $[-]$  \\
			Out-of-plane Poisson's ratio $\nu_{23}^f$                       & 0.25 $[-]$  \\
			In-plane shear modulus $G_{12}^f$, $G_{13}^f$                   & 70.0 $[GPa]$  \\
			Out-of-plane shear modulus $G_{23}^f$                           & 7.8  $[GPa]$  \\  
			Axial coefficient of thermal expansion (CTE) $\alpha_{11}^f$    & -0.54E-6 $[K^{-1}]$  \\ 
			Transverse CTE $\alpha_{22}^f$, $\alpha_{33}^f$                 & 10.08E-6 $[K^{-1}]$ \\
			Thermal conductivity $\kappa^f$                                 & 5.4 $[W/mK]$  \\
			Specific heat $c_{p}^f$                                         & 879.0 $[J/kgK]$  \\ \bottomrule
		\end{tabular} 
		\label{fiber_matprops}
	\end{table}
}  

\textbf{\begin{table}[htbp]
		\caption{Elastic and thermal material properties of the RIM R135/H1366 epoxy resin \cite{Shah_maiaru_polymers2021}. }
		\centering
		\begin{tabular}{lc}
			\toprule
			Material Property                               & Value                \\ \midrule
			Density $\rho^m $                               &  1200.0  $[kg/m^3]$  \\ 
			Elastic modulus $E^m$                           &  2482.0  $[MPa]$     \\
            Poisson's ratio $\nu^m$                         &  0.37    $[-]$     \\
			Coefficient of thermal expansion  $\alpha^m$    & 61.0E-6  $[K^{-1}]$  \\ 
			Coefficient of chemical shrinkage $\beta^m$     & 0.111 $[-]$  \\ 
			Thermal conductivity $\kappa^m$                 & 0.245    $[W/mK]$    \\
			Specific heat $c_{p}^m$                         & 1600.0   $[J/kgK]$   \\ \bottomrule
		\end{tabular} 
		\label{RIMR_epoxy_matprops}
	\end{table}
} 

\subsection{Single-element cure analysis}

\begin{figure}[htbp]
	\centering
	\includegraphics[scale=0.75]{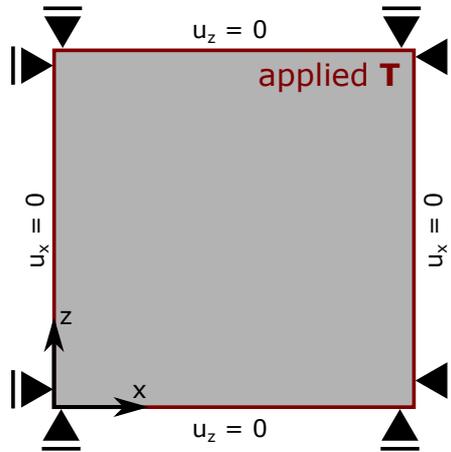}
	\caption{Schematic representation of the constrained single-element.}
	\label{SE_constrained_curing_schematic}
\end{figure}

\begin{figure}[htbp]
	\centering
	\includegraphics[scale=0.4]{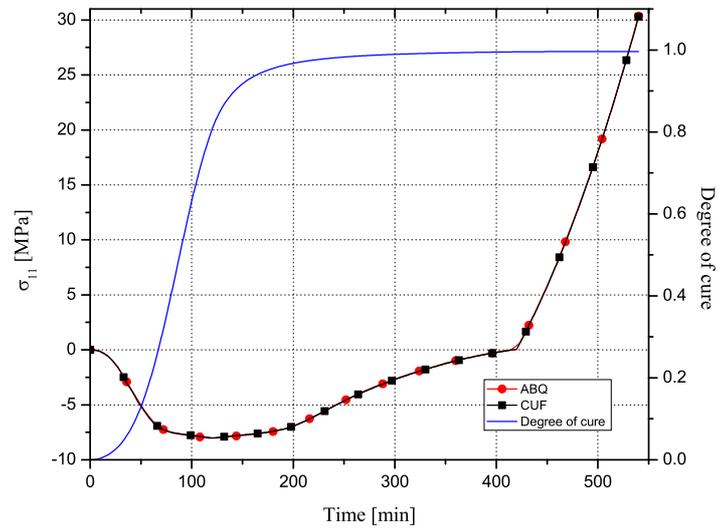}
	\caption{Evolution of residual stress (11-component) as a function of time and degree of cure for a constrained single-element.}
	\label{SE_matrix_cure_s11_dof_time}
\end{figure}

The first numerical assessment is a single-element analysis, and is a basic verification test to ensure the correctness of the implementation. The single-element is constrained on all its faces, as seen in Fig. \ref{SE_constrained_curing_schematic}, to prevent deformations due to chemical shrinkage and thermal expansion, and thereby ensures the evolution of residual stresses during the curing process. The analysis is performed using the proposed CUF approach, where the element domain is modeled with a single 4-node linear quadrilateral element (L4) on the $x-z$ face, and a single linear beam element (B2) along the $y$-axis. The analysis is also performed in Abaqus (ABQ) using a single C3D8T element, and provides a numerical reference. The evolution of the residual stress (11-component) as a function of the cure time $t$ and the degree of cure $\phi$, as predicted by the two numerical models, is plotted in Fig. \ref{SE_matrix_cure_s11_dof_time}. It is seen from the plot that the residual stress predicted by the CUF and Abaqus models are in perfect agreement with each other, and therefore provide an initial verification of the proposed numerical approach.

\subsection{Curing of square-packed RUC} \label{sec_sq_pack}

\begin{figure}[htp!]
	\begin{subfigure}[b]{0.48\textwidth}
		\centering
	\includegraphics[scale=0.5]{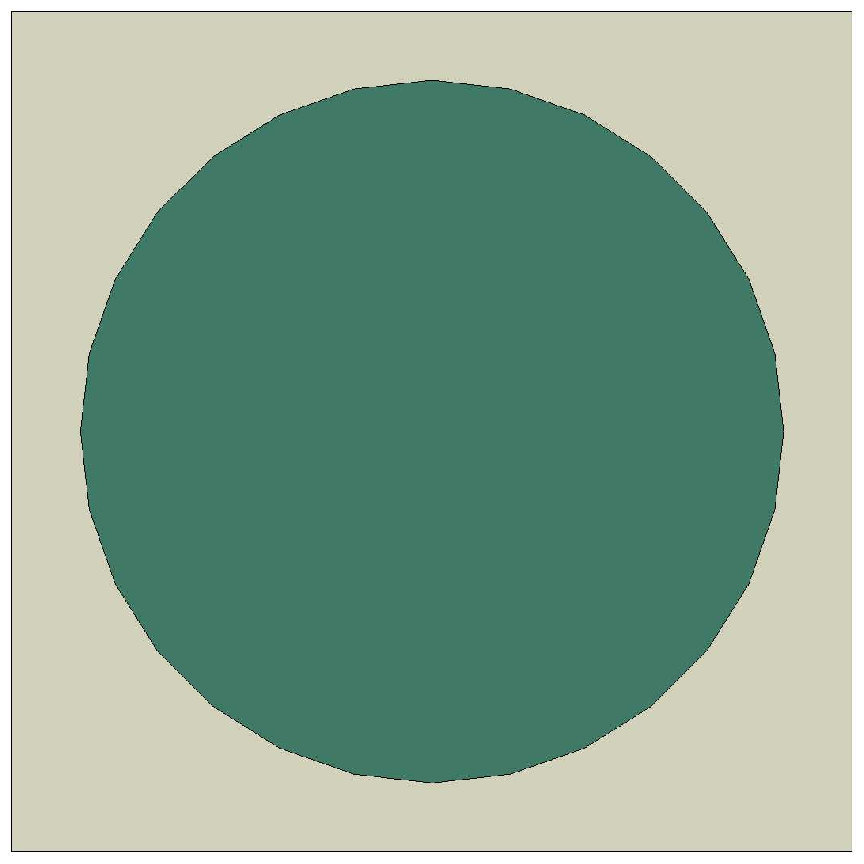}
	\caption{Square-packed RUC}
	\label{Sq_pack_RUC_schematic}
	\end{subfigure}%
	~
	\begin{subfigure}[b]{0.48\textwidth}
		\centering
    	\includegraphics[scale=0.5]{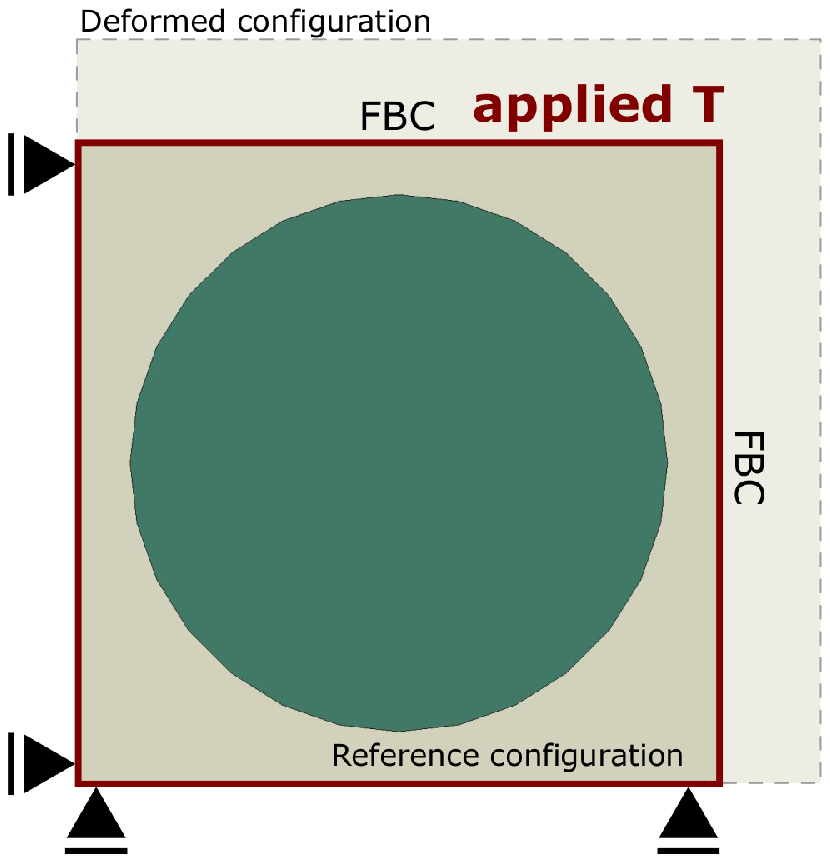}
    	\caption{Boundary conditions for curing analysis}
    	\label{Curing_BC_schematic}
	\end{subfigure}
	\caption{Schematic representation of a square-packed RUC with the applied boundary conditions.}
	\label{sq_pack_schematic_BC}
\end{figure}

The present numerical example considers the square-packed Repeating Unit Cell (RUC) with a single fiber, as shown in Fig. \ref{Sq_pack_RUC_schematic}. The boundary conditions applied on the RUC are schematically shown in Fig. \ref{Curing_BC_schematic}. A prescribed temperature based on the cure cycle (See Fig. \ref{RIMR_cure_cycle_plot}) is applied on the surface of the RUC. Flat Boundary Conditions (FBC), a special case of Periodic Boundary Conditions (PBC), are applied on the faces of the RUC which ensures that its faces remain flat in the deformed configuration. Further details on the use of FBC and its equivalence to PBC in the current application can be found in \cite{maiaru2018characterization, Shah_maiaru_polymers2021}.

\begin{figure}[htbp]
	\centering
	\includegraphics[scale=1.2]{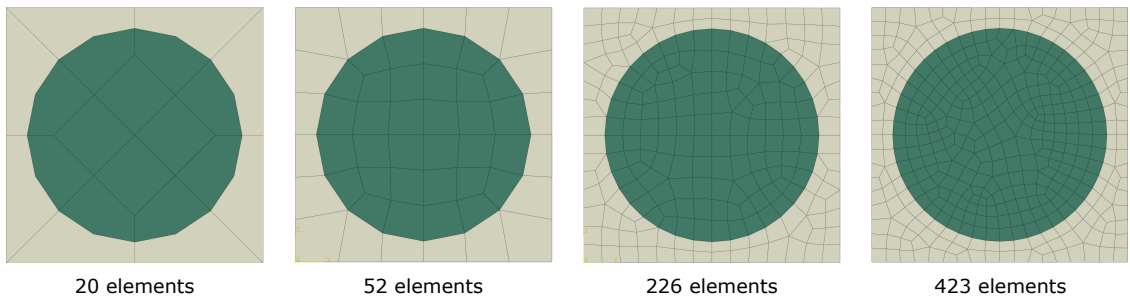}
	\caption{Meshes used in Abaqus and CUF to discretize the square-packed RUC.}
	\label{Sq_pack_all_msh}
\end{figure}

\begin{figure}[htbp]
 	\centering
 	\includegraphics[scale=0.35]{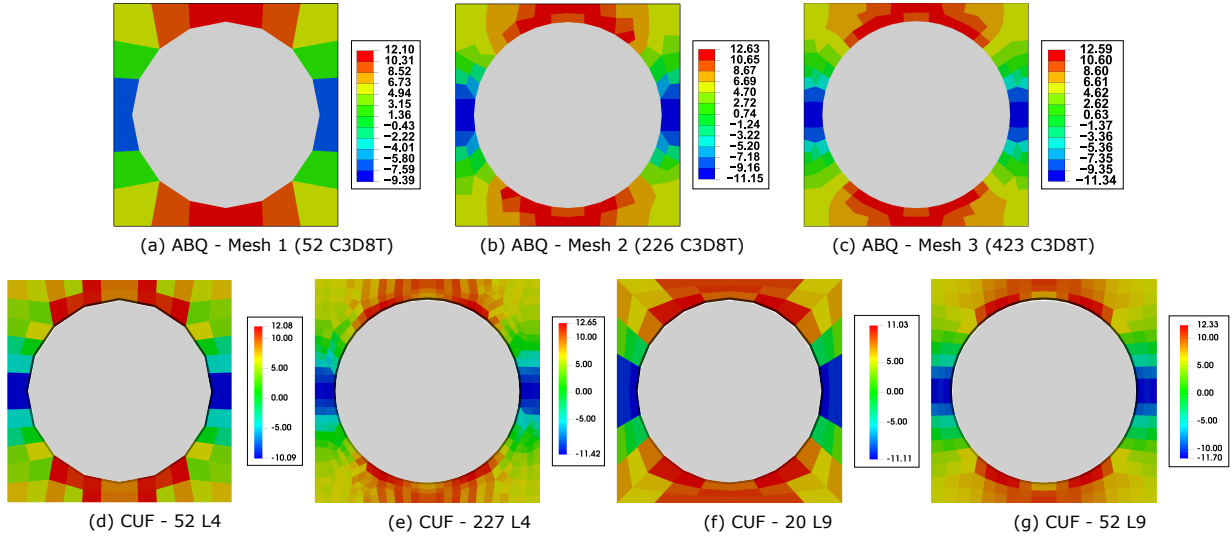}
 	\caption{Distribution of residual stress (22-component) in the square-packed RUC at the end of cure.}
 	\label{RVE_1fib_ALL_s22_contour}
 \end{figure}

The process modelling of the square-packed RUC is performed using a series of CUF models with varying levels of refinement within the RUC face, using both 4-node linear (L4) and 9-node quadratic (L9) quad elements. Each CUF model consists of a single linear beam element (B2) along the thickness direction. Two reference 3D-FE models are also developed in Abaqus, where the RUC thickness is represented by a single element. The discretization used in each numerical model is visualized in Fig. \ref{Sq_pack_all_msh}. The residual stresses in the transverse direction (22-component) that develop within the RUC at the end of the cure cycle, as predicted by the various models, is shown in the form of contour plots in Fig. \ref{RVE_1fib_ALL_s22_contour}. A summary of all the numerical models, along with the required computational time, is presented in Table \ref{sq_packed_summary_table}.

\textbf{\begin{table}[htbp]
		\caption{Summary of the numerical models used in the process simulation of the square-packed RUC. }
		\centering
			\begin{tabular}{lccc}
				\toprule
			    Model           & No. of elements & DOF   & Analysis Time {[}s{]} \\ \midrule
			    ABQ - Mesh 1    & 52 C3D8T        & 523   & 40                    \\
    		    ABQ - Mesh 2    & 226 C3D8T       & 2,043 & 52                    \\
			    ABQ - Mesh 3    & 423 C3D8T       & 3,683 & 68                    \\
			    CUF - Mesh 1    & 52 L4           & 520   & 3.2                   \\
			    CUF - Mesh 2    & 227 L4          & 2,048 & 13.4                  \\
			    CUF - Mesh 3    & 20 L9           & 712   & 4.7                   \\  
			    CUF - Mesh 4    & 52 L9           & 1,864 & 14.6                  \\ \bottomrule
		\end{tabular} 
		\label{sq_packed_summary_table}
	\end{table}
}   

From Fig. \ref{RVE_1fib_ALL_s22_contour}, it is seen that both the CUF and 3D-FE models are in very good agreement with each other, and provides further verification of the proposed numerical approach. The coarsest models in both cases, i.e. `ABQ - Mesh 1' and `CUF - 54 L4', predict very similar stress fields, and are underestimated when compared to the refined models. A similar observation is made for the case of the `CUF - 20 L9' model, where the predicted stresses are indicative of the intermediate level of refinement within the model. It is also seen that further mesh refinement, in both CUF and 3D-FE, does not lead to any significant improvements, thereby indicating mesh convergence. Comparing the associated computational time (see Table \ref{sq_packed_summary_table}) for the `ABQ - Mesh 2' analysis with those based on refined CUF (227 L4 and 52 L9), it is seen that the proposed CUF approach is approximately $4x$ as fast as the corresponding 3D-FEA, for similar levels of accuracy.

\subsection{Curing of RVE with random fiber distribution}

\begin{figure}[htbp]
	\centering
	\includegraphics[scale=0.9]{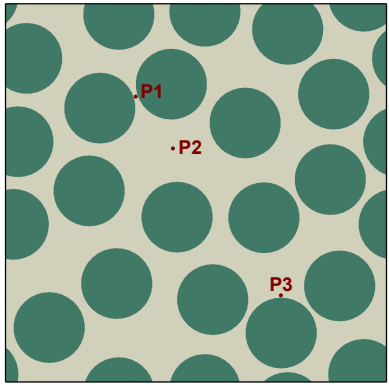}
	\caption{RVE with 20 randomly distributed fibers.}
	\label{RVE_20fib_schematic}
\end{figure}

\begin{figure}[htbp]
	\centering
	\includegraphics[scale=1.0]{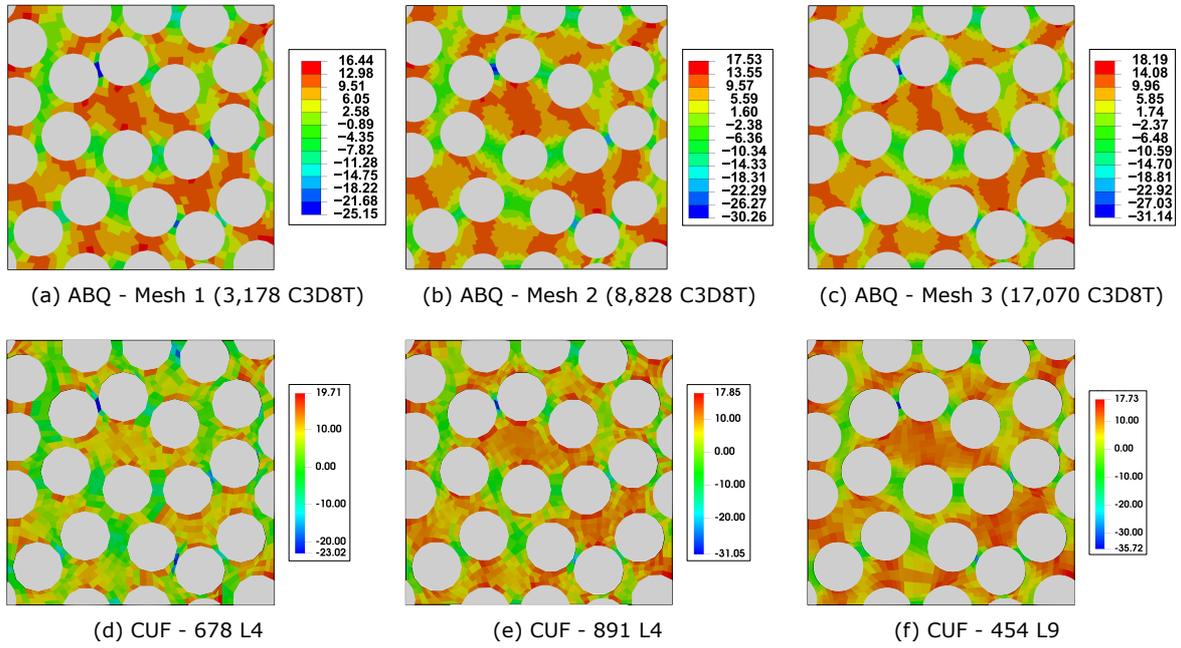}
	\caption{Distribution of residual stress (22-component) in the 20-fiber RVE at the end of cure.}
	\label{RVE_20fib_ALL_s22_contour}
\end{figure}

\begin{figure}[htp!]
	\begin{subfigure}[b]{0.32\textwidth}
		\centering
		\includegraphics[scale=0.24]{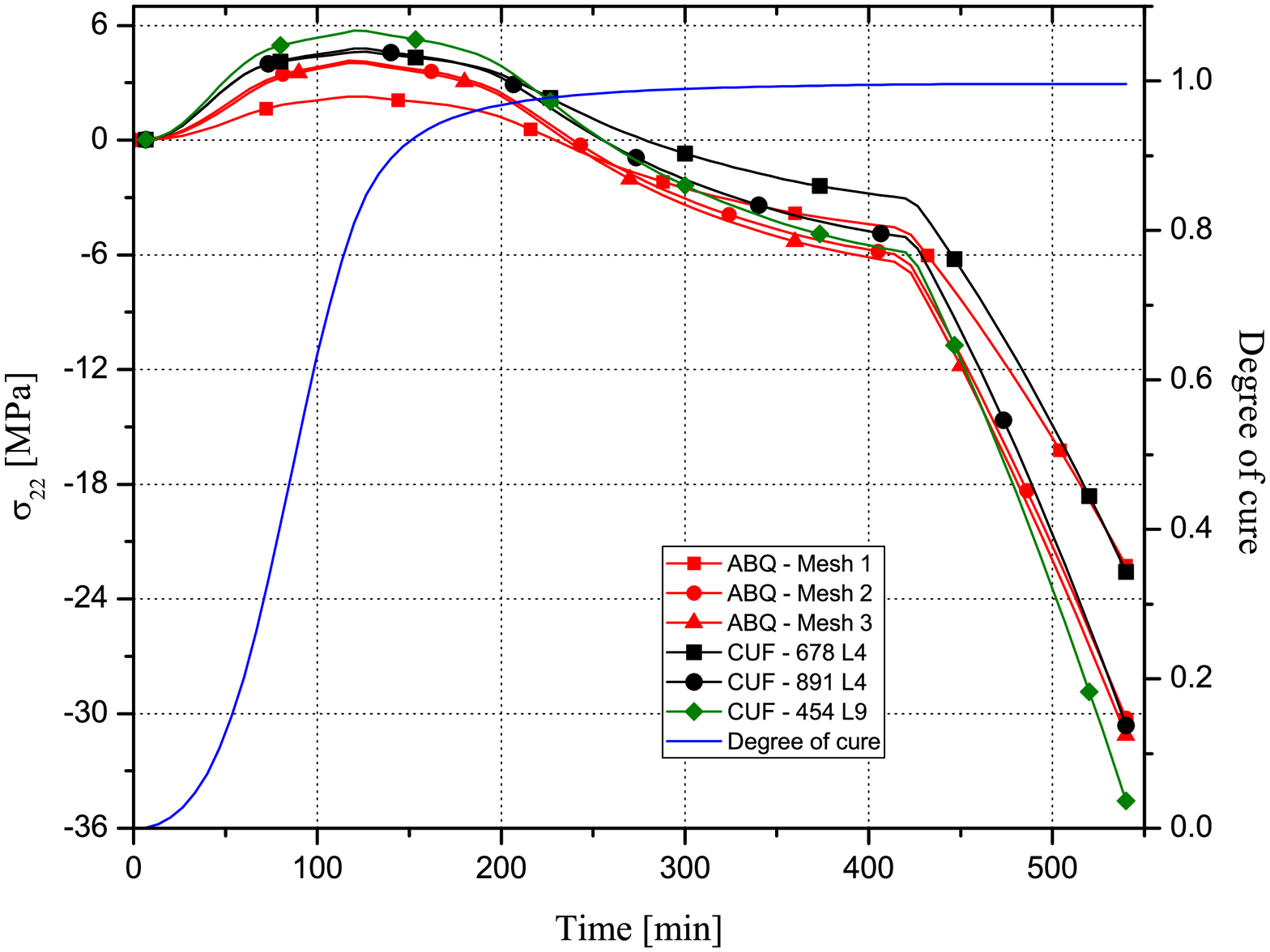}
		\caption{P1}
		\label{RVE_20fib_S22_min}
	\end{subfigure}%
	~
	\begin{subfigure}[b]{0.32\textwidth}
		\centering
		\includegraphics[scale=0.24]{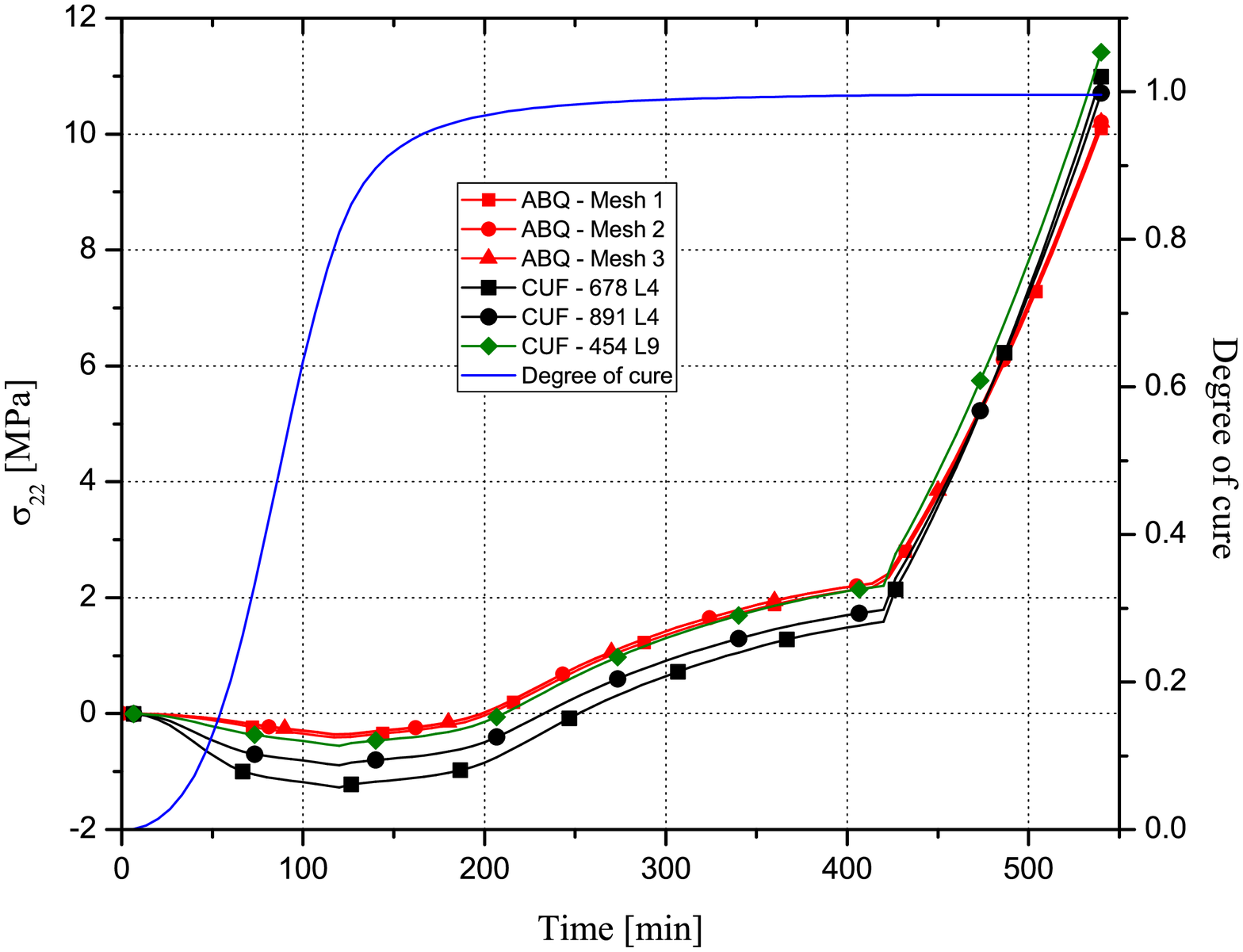}
		\caption{P2}
		\label{RVE_20fib_S22_centre}
	\end{subfigure}
	~
	\begin{subfigure}[b]{0.32\textwidth}
		\centering
		\includegraphics[scale=0.24]{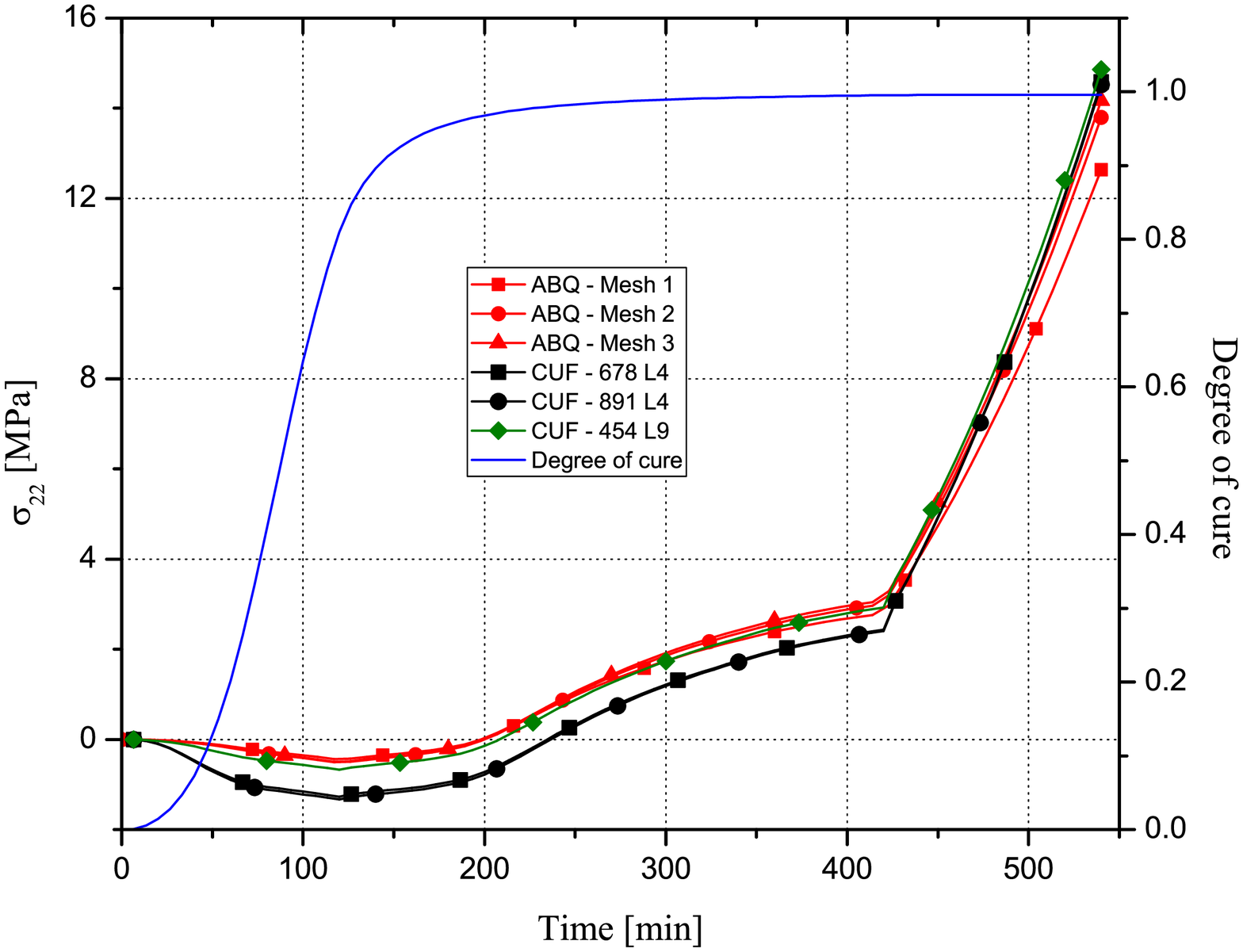}
		\caption{P3}
		\label{RVE_20fib_S22_P3}
	\end{subfigure}
	\caption{Evolution of residual stress (22-component) as a function of cure time.}
	\label{RVE_20fib_S22_time_plot}
\end{figure}

This assessment considers a periodic RVE with 20 randomly distributed fibers, as shown in Fig. \ref{RVE_20fib_schematic}. The RVE boundary conditions described in Section \ref{sec_sq_pack} are applied in the current analysis. A series of CUF models is developed with increasing levels of refinement within the RVE face, and a single linear beam element (B2) is used to represent the RVE thickness in each model. Three 3D-FE models are also developed in Abaqus as a numerical reference. The residual stress (22-component) predicted by the models at the end of the cure cycle is presented in Fig. \ref{RVE_20fib_ALL_s22_contour}. The residual stress evolution at three specific points within the RVE (see Fig. \ref{RVE_20fib_schematic}), as a function of cure time, is plotted in Fig. \ref{RVE_20fib_S22_time_plot}. A summary of the computational models is presented in Table \ref{RVE_20fib_summary_table}.

From Fig. \ref{RVE_20fib_ALL_s22_contour}, it is seen that successive refinement of the mesh leads to a converged solution in both the 3D-FE and CUF models. The coarsest models significantly underestimate the compressive stresses that develop at the point P1, which is the matrix region between two neighboring fibers, and thus a zone of considerable stress concentration within the RVE. This can be observed in Fig. \ref{RVE_20fib_S22_min}, where the `ABQ - Mesh 1' and `CUF - 678 L4' models both predict similar magnitudes of the developed residual stress, which is in strong contrast to that predicted by the more refined models. On the other hand, considering the stress evolution at the point P2 (see Fig. \ref{RVE_20fib_S22_centre}), it is seen that all the model predictions are in good general agreement. This is attributed to the fact that the point P2 is situated in a matrix-rich region that does not involve thermal gradients or stress concentrators, and therefore a lower mesh density is sufficient. Finally, examining the residual stress evolution at the point P3 (see Fig. \ref{RVE_20fib_S22_P3}), it is seen that the coarsest 3D-FE model, i.e. `ABQ - Mesh 1', underestimates the post-cure stress magnitude, and while this is not as inaccurate as in the case of Point P1, still has a considerable error with respect to the refined models. This is explained by the fact that the Point P3 lies in the immediate vicinity of a single fiber which acts as a stress concentrator. The trends observed in the behavior of the `ABQ - Mesh 1' model at the points P1, P2 and P3 is therefore consistent with the level of stress concentration experienced by the matrix at these points. It is noted that the corresponding coarsest CUF model, i.e. `CUF - 678 L4', predicts a post-cure residual stress which is in good agreement with that reported by refined models, at the Point P3, inferring that the coarsest CUF model performs better than the corresponding 3D-FE model. 

Considering the significant variation in stress concentration at different points within the RVE, which is a consequence of the randomly distributed fibers, any numerical model would require a refined discretization of the matrix component in order to accurately predict the post-cure residual stresses. The `ABQ - Mesh 2' model is the coarsest 3D-FE model whose results are sufficiently accurate, based on the mesh convergence study. The corresponding CUF model, with an equivalent quality of predicted results, is the `CUF - 891 L4' model. Comparing the computational costs associated with these two models (see Table \ref{RVE_20fib_summary_table}) it is seen that the CUF approach is approximately $10x$ as fast as the 3D-FE case, and is over $7x$ smaller in computational size based on the number of DOF within the models. An important observation is that the computational efficiency of CUF over 3D-FE, when the model domain is increased from a single-fiber RUC to a 20-fiber RVE, correspondingly increases from approximately $4x$ to $10x$, indicating the superior scalability of CUF. Finally, the `CUF - 454 L9' model, while very accurate, has an unnecessarily excessive level of kinematic refinement, and the corresponding higher computational cost therefore implies that a sufficiently refined L4 model (such as the `CUF - 891 L4' model) is preferable over those based on L9, for the current class of problem. It is, however, noted that even the `CUF - 454 L9' model is about $3.5x$ as fast as the most refined 3D-FE model, i.e. `ABQ - Mesh 3', and is $6.8x$ smaller in size, demonstrating the computational efficiency of the CUF approach over conventional 3D-FEA.

\textbf{\begin{table}[htbp]
		\caption{Summary of the numerical models used in the process simulation of the 20-fiber RVE. }
		\centering
		\begin{tabular}{lccc}
			\toprule
			Model           & No. of elements &   DOF    & Analysis Time {[}s{]} \\ \midrule
			ABQ - Mesh 1    &  3,178 C3D8T    &  19,911  & 352                  \\
			ABQ - Mesh 2    & 8,828 C3D8T     & 54,447   & 830                   \\
			ABQ - Mesh 3    & 17,070 C3D8T    & 104,403  & 1622                   \\
			CUF - Mesh 1    & 678 L4          & 5,728    & 55                   \\
			CUF - Mesh 2    & 891 L4          & 7,432    & 85                 \\
			CUF - Mesh 3    & 454 L9          & 15,400   & 463                 \\ \bottomrule
		\end{tabular} 
		\label{RVE_20fib_summary_table}
	\end{table}
}  

\section{Conclusion} \label{sec_conclusions}

A novel computational process modeling framework, based on CUF-derived higher-order finite elements, has been proposed for the curing analysis of thermoset fiber-reinforced composites at the micro-scale, and the prediction of residual stresses that develop during the cure cycle. A description of the thermoset process model and CUF structural modeling is presented. A set of numerical assessments has been considered as verification cases, where the CUF approach is compared with reference 3D finite element models developed in Abaqus. The first assessment is the curing analysis of a single-element, and the perfect agreement between CUF and 3D-FE provides an initial verification of the implementation. Subsequent assessments consider a single-fiber repeating unit cell and a 20-fiber representative volume element, and the accuracy of the CUF models is demonstrated. It is also shown that CUF models exhibit an order-of-magnitude higher computational efficiency -- in terms of analysis time -- than 3D-FE models, for comparable levels of accuracy. By comparing the computational costs of the two numerical methods for the single-fiber RUC and 20-fiber RVE cases, it is seen that CUF scales better than 3D-FEA as the structural size is increased. Such advantages, inherent to CUF, make it a strong candidate in applications such as virtual manufacturing and testing with a multiscale resolution, which fall under the scope of Integrated Computational Materials Engineering (ICME) \cite{nasa_2040_vision}.

\section{Acknowledgments}
This material is based upon work supported by the National Science Foundation under Award \#2145387. Any opinions, findings, and conclusions or recommendations expressed in this material are those of the author(s) and do not necessarily reflect the views of the National Science Foundation.

\clearpage
\bibliographystyle{unsrt}
\bibliography{bib_cuf_curing}

\end{document}